\documentclass[reprint,amsmath,amssymb,aps,prb]{revtex4-1}
\usepackage{amsmath,amssymb}
\usepackage{graphicx}
\usepackage{epsfig}
\usepackage{psfrag}
\usepackage{hyperref}
\usepackage{placeins}
\usepackage{float}
\usepackage{xcolor}
\usepackage{bm}

\newcommand{\mathsym}[1]{{}}
\newcommand{\unicode}[1]{{}}

\begin{document}

\title{Photo-Induced Anomalous Hall Effect in Two-Dimensional Transition-Metal
  Dichalcogenides}

\author{Phuong X. Nguyen}
\author{Wang-Kong Tse}
\affiliation{
   Department of Physics and Astronomy, The University of Alabama, Alabama 35487, USA
}

\date{\today}

\begin{abstract}
  A circularly polarized \textit{a.c.} pump field illuminated near resonance on two-dimensional transition metal dichalcogenides (TMDs) produces an anomalous Hall effect in response to a \textit{d.c.} bias field.  In this work, we develop a theory for this photo-induced anomalous Hall effect in undoped TMDs irradiated by a strong coherent laser field. The strong field renormalizes the equilibrium bands and opens up a dynamical energy gap where single-photon resonance occurs. The resulting photon dressed states, or Floquet states, are treated within the rotating wave approximation. 
  A quantum kinetic equation approach is developed to study the  non-equilibrium density matrix and time-averaged transport currents under the simultaneous influence of the strong \textit{a.c.} pump field and the weak \textit{d.c.} probe field. Dissipative effects are taken into account in the kinetic equation that captures relaxation and dephasing. 
  The photo-induced longitudinal and Hall conductivities display notable resonant signatures when the pump field frequency reaches the spin-split interband transition  energies. Rather than valley polarization, we find that the anomalous Hall current is mainly driven by the intraband response of photon-dressed electron populations near the dynamical gap at both valleys,
accompanied by a smaller contribution due to interband coherences. 
%  and to a lesser extent by an asymmetric distribution of the interband coherences. 
%  we find that the anomalous Hall current is mainly driven by an asymmetric momentum-space distribution of photon-dressed electron population near the dynamical gap, and to a lesser extent by an asymmetric distribution of the interband coherences. 
  These findings highlight the importance of photon-dressed bands and non-equilibrium distribution functions 
  in achieving a proper understanding of photo-induced anomalous Hall effect in a strong pump field.
\end{abstract}

\maketitle

\section{Introduction} \label{sec:intro}

Since the discovery of graphene \cite{novoselov2004electric}, van der Waals materials have emerged as a broad family of two-dimensional (2D) layered materials with diverse physical properties ranging from semimetals, semiconductors, insulators to 2D ferromagnets and superconductors \cite{novoselov20162d}.
%emerges, namely van der Waals materials, with diverse physical properties, from semimetal, semiconductor, insulator to 2D magnet and superconductor \cite{novoselov20162d}.
Two-dimensional transition metal dichalcogenides (TMDs) (\textit{e.g.} MoS$_2$, WS$_2$, MoSe$_2$, WSe$_2$) are van der Waals semiconductors with a band gap within the visible spectrum. In monolayers, TMDs exhibit broken spatial inversion symmetry combined with strong spin-orbit interaction, resulting in a large valence band splitting 
appearing across the direct gaps at the valleys K and K' \cite{zhu2011giant} with inherently coupled spin and valley degrees of freedom  \cite{xiao2012coupled}. Through the valley selection rule, carriers near the valence band edge at each of the valleys couple preferentially to light with a definite circular polarization, allowing them to be selectively excited to the conduction band. 
For frequencies above the band gap, the optical excitation creates a carrier population imbalance between the two valleys, \textit{i.e.} a valley polarization \cite{cao2012valley}.

If the system is additionally driven by a \textit{d.c.} electric field, valley-resolved photovoltaic transport occurs. In particular, an anomalous Hall effect will result from the net transverse charge current due to  unbalanced population of  photoexcited K and K' carriers \cite{valley_Niu_1,valley_Niu_2}. 
A similar Hall effect, caused by photo-induced spin polarization, has also been predicted \cite{AHE_Zhang} in semiconductor systems due to spin-orbit coupling and observed \cite{AHE_Exp1, AHE_Exp2, AHE_Exp3,AHE_Exp4} in III-V semiconductor structures, Bi$_2$Se$_3$ topological insulator thin film \cite{AHE_Exp5} and few-layer WTe$_2$ Weyl semimetal \cite{AHE_Exp6}. In TMDs, this photo-induced anomalous Hall effect has been recently observed in illuminated samples of monolayer MoS$_2$ as well as bilayer MoS$_2$ placed under an  out-of-plane electric field \cite{mak2014valley,lee2016electrical}. It  has also been recently observed in illuminated samples of exfoliated graphene \cite{AHE_Exp7,AHE_Exp8}, in which the Hall effect is purely due to optically-induced Berry curvature.

Early theoretical treatments on the photo-induced Hall effect in TMDs have %so far
been largely focused on the role of valley selection rules and Berry curvatures obtained from the equilibrium bands, with  the tacit assumption that the optical excitation is sufficiently weak that the electronic band structure remains unaltered under irradiation. Hall transport in the regime of strong optical excitations, which can reveal rich quantum dynamics through photon dressing effects and are readily realizable in experiments, has received increasing theoretical attention \cite{Oka1,Torres,AHE_Mitra,AHE_Ran,AHE_Law,WRL1}.
In a recent experiment \cite{sie2015valley}, dynamic Stark shift of the bands has been observed in WS$_2$ under a strong optical pump field with subgap frequency. When the pump frequency is above the band gap, hybridization between the photon-dressed valence and conduction 
bands generates a dynamical gap \cite{Galitskii_EM,OptStark1}. The hybridized states, which are also known as Floquet states, have not yet been observed in TMDs but has been directly observed in topological insulator surface states \cite{GedikFloquet1, GedikFloquet2}. The realization of Floquet states provides a means to realize many interesting non-equilibrium phenomena \cite{Oka_Review}, such as  Floquet topological phases \cite{lindner2011floquet, Floq_Phase1}, Floquet control of exchange interaction \cite{Floq_RKKY1,Floq_RKKY2} and tunneling \cite{Floq_Tunnel}, and Floquet time crystals \cite{Floq_TC1}. 

Under strong optical excitation by the pump field, the valley-resolved Hall effect is influenced by the  photon renormalization of the electronic bands as well as non-equilibrium carrier kinetics \cite{kovalev2018valley}. In this work, we provide a density matrix formulation for photo-induced valley Hall transport that allows us to treat the photon-dressed bands and carrier kinetics in a single framework. Our theory is developed using the rotating wave approximation, which provides better analytic insights compared to full numerical solutions, in the regime of near resonance and weak coupling where multiphoton effects are unimportant \cite{WRL1}. Band populations and interband coherences are obtained in a transparent manner from the solution of the kinetic equation of the density matrix. These are then used to compute the photo-induced valley polarization and longitudinal and Hall photoconductivities. Our findings reveal that the physical picture behind the photo-induced anomalous Hall effect is much more nuanced in a strong laser field than the commonly assumed picture of valley population imbalance, due to the formation of different photon-dressed bands at the two valleys. 

Our paper is organized as follows. Sec.~\ref{sec:Ham} lays out the model of our system and discusses the photon-induced renormalization of the equilibrium band structure. We then introduce the density matrix formalism and the kinetic equation governing its dynamics in Sec.~\ref{sec:QKE}. In Secs.~\ref{sec:p0th}-\ref{sec:p1st}, we solve the kinetic equation and obtain the density matrix of the pumped system, first in the absence and then in the presence of the \textit{d.c.} electric field. Sec.~\ref{sec:dcI} then presents the derivation of the photovoltaic longitudinal and Hall currents and our numerical results of the photoconductivities, followed by conclusion in Sec.~\ref{sec:concl}. 

\section{Model of TMD Coupled to Optical Pump Field} \label{sec:Ham}

The low-energy Hamiltonian of 2D TMD is given by\cite{xiao2012coupled}
\begin{eqnarray} \label{orgham}
H_0 & =& v(\tau  k_x\sigma_x + k_y\sigma_y) +(\hat{\Delta} - \frac{\lambda \tau}{2}s_z) \sigma_z + \frac{\lambda \tau}{2} s_z, \quad \label{h0}
\end{eqnarray}
where  $\bm{\sigma}$ denotes the vector of Pauli matrices, $2 \hat{\Delta}$ is the band gap energy, $2\lambda$ is the spin-orbit splitting of the valence bands, $\tau = \pm 1$ is the valley index for K and K' respectively and $s_z=\pm 1 $ the spin index for up and down. In the vicinity of each valley, the low-energy physics is described by two copies of spin-resolved Dirac Hamiltonian with band gap $2\Delta_{1,2} = 2\hat{\Delta}\mp {\lambda}$. In this paper, we take MoS$_2$ as the prototypical example of TMDs and use the corresponding values \cite{xiao2012coupled} of band gap $2 \hat{\Delta} = 1.66$ eV and spin-orbit splitting $2\lambda = 0.15$ eV  for our numerical calculations. 

We can develop our theory for one spin $s$ and one valley $\tau$ and obtain the total photovoltaic current at the end by summing the contributions from both spins and both valleys. Dropping the inessential energy shift from the last term, Eq.\eqref{h0} takes the typical form of a massive Dirac Hamiltonian 
\begin{eqnarray}
H_0 & =& v(\tau  k_x\sigma_x + k_y\sigma_y) +\Delta \sigma_z, \label{h0sim}
\end{eqnarray}
here $\Delta = \hat{\Delta}-\tau s_z{\lambda}/2$, which takes the two values $\Delta_{1,2}$ corresponding to $\tau s_z = \pm 1$. 
Diagonalizing Eq.\eqref{h0sim} gives the conduction ($+$) and valence ($-$) band energy $\pm \alpha_k = \pm \sqrt{(v k)^2+\Delta^2}$ and the corresponding spinor wavefunctions,
\begin{eqnarray} \label{spinors}
\chi_{k+}=\begin{bmatrix} \cos({\theta_k}/{2}) \\ \sin({\theta_k}/{2}) e^{i \tau \phi} \end{bmatrix},\,\, \chi_{k-}=\begin{bmatrix} -\sin({\theta_k}/{2}) \\ \cos({\theta_k}/{2}) e^{i \tau \phi} \end{bmatrix},
\end{eqnarray}
where we have defined $\cos\theta_k =\Delta/\alpha_k$, $\sin\theta_k=\tau v k/\alpha_k$, and $\tan\phi=k_y/k_x$. 

The pump field laser is described by an \textit{a.c.} electric field $\bm{E} = E_0(\cos\omega t\hat{x}+\mu\sin\omega t\hat{y})$, in which $\mu=\pm1$ denotes the left and right circular polarization. The light-matter interaction Hamiltonian follows from the minimal coupling $\bm{k} \to \bm{k}+e\bm{A}$ (where $e > 0$ is the electronic charge) with the vector potential $\bm{A}=-\int\bm{E}dt=-(E_0/\omega)(\sin\omega t \hat{x}-\mu\cos\omega t\hat{y})$. The total Hamiltonian then becomes $H = H_0-(\Lambda/2)(\tau\sin\omega t\sigma_x-\mu\cos\omega t\sigma_y)$, where $\Lambda=2eE_0v/\omega$. The pump field couples to the orbital degrees of freedom only and optical transitions preserve spins. 

\begin{figure}[hp!tb]
  \includegraphics[width=8.5cm,angle=0]{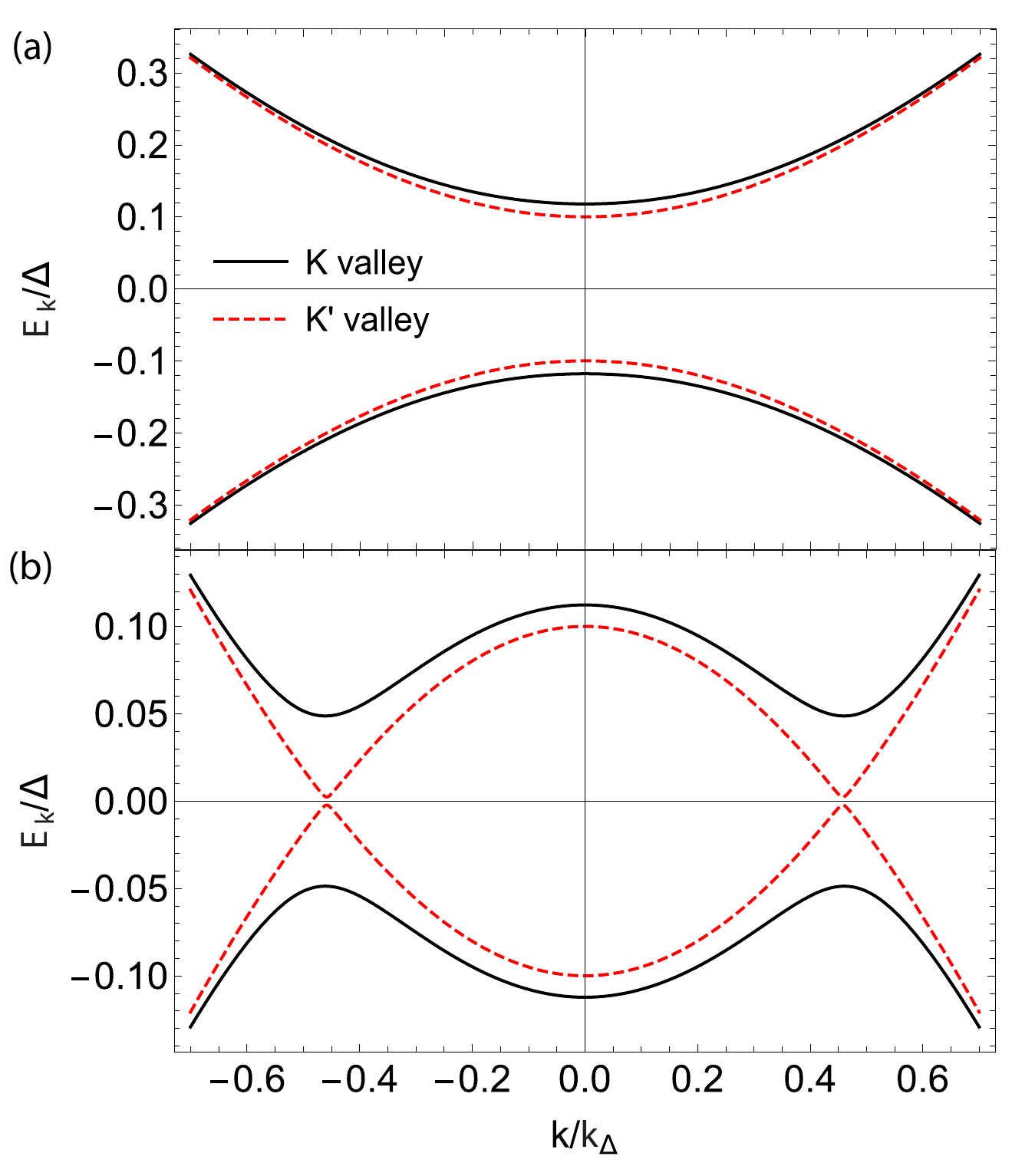}
    %band_combined_axes.pdf}
  \caption{(Color online). Photon-dressed bands $\pm\tilde{E}_k$ for spin-up electrons at valley K (solid) and spin-down electrons at valley K' (dashed) under a circularly polarized pump field with strength $E_0= 200$\; MV/m and helicity $\mu = 1$. The equilibrium band gap value is taken as $2\Delta = 2\Delta_1=1.585$ eV. (a) corresponds to sub-gap pump field frequency $\omega=1.4$ eV and (b) to above-gap frequency $\omega= 1.74$ eV. Energy values are scaled by $\Delta_1$ and momentum $k_x, k_y$ by $k_{\Delta_1}=\Delta_1/v$.} \label{Ek1}
\end{figure}

It will be convenient to define \cite{Dimi1,Dimi2,tse2016coherent} a set of mutually perpendicular pseudospin unit vectors $\{\hat{\boldsymbol{\alpha}}_{k}, 
\hat{\boldsymbol{\beta}}_{k},
\hat{\boldsymbol{\gamma}}_{k}\}$ and corresponding basis matrices $(\sigma_\alpha,\sigma_\beta,\sigma_\gamma)=\bm{\sigma} \cdot (\hat{\bm{\alpha}}_k,\hat{\bm{\beta}}_k,\hat{\bm{\gamma}}_k)$ to rewrite the Hamiltonian. With the definition  
$\hat{\boldsymbol{\kappa}}_\tau \equiv \cos\phi\hat{\bm{x}}+\tau\sin\phi\hat{\bm{y}}$, we define the unit vectors as
\begin{eqnarray}
\hat{\boldsymbol{\alpha}}_{k} & = & \sin\theta_k \hat{\boldsymbol{\kappa}}_\tau+\cos\theta_k\hat{\bm{z}}, \\
\hat{\boldsymbol{\beta}}_{k} & = & \tau \hat{\bm{z}}\times\hat{\boldsymbol{\kappa}}_\tau, \\ 
\hat{\boldsymbol{\gamma}}_{k}& = & -\tau\cos\theta_k\hat{\boldsymbol{\kappa}}_\tau+\tau\sin\theta_k\hat{\bm{z}}. 
\end{eqnarray}
$\{\hat{\boldsymbol{\alpha}}_{k}, \hat{\boldsymbol{\beta}}_{k}, \hat{\boldsymbol{\gamma}}_{k}\}$ forms a right-handed triad defined locally at each $\bm{k}$ point. Note that they are dependent on the valley index $\tau$. $\{\sigma_\alpha,\sigma_\beta,\sigma_\gamma\}$ are the corresponding pseudospin projections
\begin{eqnarray}
\sigma_\alpha &=& \begin{bmatrix} \cos\theta_k & \sin\theta_k e^{-i\tau\phi} \\ \sin\theta_k e^{i\tau\phi} & -\cos\theta_k  \end{bmatrix}, \\
\sigma_\beta &=& i\tau \begin{bmatrix} 0 & -e^{-i\tau\phi} \\ e^{i\tau\phi} & 0  \end{bmatrix}, \\
\sigma_\gamma &=& \tau\begin{bmatrix} \sin\theta_k & -\cos\theta_k e^{-i\tau\phi} \\  -\cos\theta_k e^{i\tau\phi} & -\sin\theta_k \end{bmatrix},
\end{eqnarray}
 It is also useful to note that $\sigma_{\alpha,\beta,\gamma}$ is related to the usual Pauli matrices $\sigma_{x,y,z}$ through the pseudospin-to-band unitary transformation $\mathcal{U}_k \equiv [\chi_{k+}\, \chi_{k-}]$ by 
$\sigma_{\alpha} = \mathcal{U}_k\sigma_z\mathcal{U}^{\dagger}_k$, $\sigma_{\beta} = \mathcal{U}_k(\tau\sigma_y)\mathcal{U}^{\dagger}_k$, $\sigma_{\gamma} = \mathcal{U}_k(-\tau\sigma_x)\mathcal{U}^{\dagger}_k$. We can then represent the total Hamiltonian as follows, 
\begin{eqnarray}
H &=& \left[\alpha_k +\frac{\Lambda}{2}\mu\tau\sin\theta_k\sin(\phi-\mu\omega t)\right]\sigma_\alpha \label{totalHam} \\
&&+\frac{\Lambda}{2}\mu\tau\cos(\phi-\mu\omega t)\sigma_\beta -\frac{\Lambda}{2}\mu\cos\theta_k\sin(\phi-\mu\omega t)\sigma_\gamma.  \nonumber
\end{eqnarray}
The total Hamiltonian above, now expressed in the new pseudospin representation, can be further transformed into the rotating frame using the unitary transformation $U=e^{-i\omega t\sigma_\alpha/2}$ as  $\tilde{H} = U^{\dagger}HU - i U^{\dagger}{\partial_t U}$. Hereafter, quantities in the rotating frame will be denoted with an overhead tilde.  In the rotating wave approximation (RWA), we retain only time-independent terms and obtain the rotating-frame Hamiltonian as 
\begin{eqnarray}
\tilde{H}  
= \left(\alpha_k-\frac{\omega}{2}\right)\sigma_\alpha +\frac{\Lambda}{4}M_k\left(\cos\phi \sigma_\beta-\mu\sin\phi\sigma_\gamma\right),  \label{ham'}
\end{eqnarray}
where $M_k \equiv \mu\tau+\cos\theta_k$ captures the valley selection rule at $k = 0$ with $M_k = 2$ when $\mu = \tau$, and zero otherwise. Diagonalizing the Hamilonian $\tilde{H}$ gives the photon-dressed conduction and valence band dispersions in the rotating frame,
\begin{eqnarray} \label{eq:banddisper}
\tilde{E_k}=\pm \sqrt{\left(\alpha_k-\frac{\omega}{2}\right)^2+\left(\frac{\Lambda}{4}M_k\right)^2}. 
\end{eqnarray}
Fig.~\ref{Ek1} shows the photon-dressed bands of the  spin-up electrons at valley K and the spin-down electrons at valley K' for the cases when the light frequency is below and above the band gap. For circularly polarized light, the dispersion $\tilde{E_k}$ is isotropic in the $k$-space since $M_k$ is independent of $\phi$.
For the case of subgap frequency $\omega <2\Delta$ in Fig.~\ref{Ek1}(a), the band gap is enhanced from the equilibrium value due to dynamical Stark effect \cite{haug2009quantum}, becoming $\sqrt{\delta_{\mathrm{d}}^2+{\Lambda}^2}$ in the  rotating frame where $\delta_{\mathrm{d}} =  2\Delta-{\omega}$ is the detuning. One notices that the difference between the photon-dressed bands at the two valleys is quite small even at large fields. A more dramatic difference can be seen when the frequency exceeds the band gap in Fig.~\ref{Ek1}(b). At both valleys, a dynamical gap is opened at a  finite $k$ value. The gap is sizeable $\sim 77.2\,\mathrm{meV}$ at valley K but is minuscule $\sim 3.7\,\mathrm{meV}$ at valley K', which can be barely resolved at the scale of the plot. 

The drastic difference between the two photon-dressed bands is a result of the valley-selective coupling of electrons with circularly polarized light through the matrix element $M_k$. From Eq.~\eqref{eq:banddisper}, the magnitude of the gap can be found as $\Lambda M_{k = k_{\rm r}}/2$, where $k_{\rm r}$ is the momentum at which resonance transition occurs when  $2\alpha_{k_{\rm r}} = \omega$. For frequency values near the TMD band gap such as $\omega = 1.7\,\mathrm{eV}$, to generate a {dynamical gap of  $10-100\,\mathrm{meV}$} at valley K, the range of \textit{a.c.} field amplitude required is $25-250$ MV/m, which is attainable in state-of-the-art ultrafast optical experiments \cite{sie2015valley,sim2016selectively,sie2017large}. In free-standing graphene, a strong circularly polarized light similarly opens up a dynamical gap at the Dirac points, and in recent experiments the induced Dirac gap is estimated to be $\sim 70\,\mathrm{meV}$ \cite{AHE_Exp7}.

The photon-dressed states Eq.~(\ref{eq:banddisper}), which are obtained within RWA, capture similar physics as the Floquet states in the $2\times 2$ truncated Floquet space in the neighborhood of a dynamical gap \cite{Floq_Band1,Floq_Band2,Floq_Band3}, with $\pm$ in Eq.~(\ref{eq:banddisper}) corresponding to the Floquet quasienergies of the $0^{\mathrm{th}}$ conduction and $1^{\mathrm{st}}$ valence sidebands. For near-resonance frequencies $\omega \approx 2\Delta_1$ in TMDs, the dimensionless light-matter coupling parameter $\lambda = eE_0 v/(\hbar\omega^2) \lesssim 10^{-2} \ll 1$ for $E_0$ up to $250$ MV/m, therefore the system is well within the  weak coupling (also known as weak drive) regime in which RWA is expected to provide an excellent approximation. 

\section{Kinetic equation} \label{sec:QKE}

In order to calculate the photocurrent response, we first obtain the density matrix $\rho_k$  in the presence of the pump and probe fields. The Hamiltonian $H$ including the pump field vector potential is treated as the non-perturbative part of the problem. The perturbative part is due to the weak \textit{d.c.} probe field $\bm{E}$, which is  included in the Hamiltonian in the form of a slowly-varying scalar potential $\Phi(\bm{r})$ such that  $e\bm{E} = \nabla\Phi(\bm{r})$. We follow the standard procedure to derive the equation of motion for the one-time density matrix using the non-equilibrium Green's function formalism \cite{Jauho_book,Rammer}. After obtaining the quantum kinetic equation of the two-time lesser Green's function $G^<$, performing the Wigner transformation and gradient expansion, the equation of motion for the density matrix can be obtained from the kinetic equation of $G^<$ in the equal-time limit, which in frequency space translates to the following relation 
\begin{equation}
\rho_{{k}}(t) = -i\int_{-\infty}^{\infty}\frac{d\omega}{2\pi} G_{{k,\omega}}^<(t). \label{DMG} 
\end{equation}
Performing the above steps we then find the kinetic equation for $\rho_k$:
\begin{eqnarray}
\frac{\partial \rho_k}{\partial t}-e\bm{E}\cdot\frac{\partial \rho_k}{\partial \bm{k}}+i\left[H,\rho_k\right] &=& I_{\mathrm{s}}[\rho_k], \label{QKEE}
\end{eqnarray}
where $H$ is the total Hamiltonian including the optical pump field in Sec.~\ref{sec:Ham}, and $ I_{\mathrm{s}}[\rho_k]$ respresents the scattering integral that describes damping effects of relaxation and dephasing. Here intraband drift motion due to the \textit{d.c.} field is included via the second term on the left hand side of the kinetic equation. Since $\rho_{{k}}$ is a $2\times 2$ density matrix in the pseudospin space, it can be decomposed using the basis 
$\left\{\mathbb{I},\sigma_\alpha,\sigma_\beta,\sigma_\gamma\right\}$ as 
\begin{equation}
\rho_k=n_k\mathbb{I}+\frac{1}{2}\bm{S}_k\cdot\bm{\sigma}. \label{rhoDM}
\end{equation}
$n_k$ and $\bm{S}_k$ have the meanings of a charge and a pseudospin distribution function, respectively. In this work, we confine ourselves to considering carrier scattering processes that are spin-conserving and valley-conserving. This assumption is valid when no magnetic impurity is present and atomic-scaled defects that give rise to intervalley scattering are negligible. Our approach here can be easily extended to include scattering that flips spins and valleys \cite{spin_noise}. Then, in the relaxation time approximation \cite{haug2009quantum}, the scattering integral takes the following form with phenomenological  longitudinal relaxation rate $\Gamma$ and transverse relaxation rate $\Gamma_\perp$,
\begin{eqnarray}
I_{\mathrm{s}}[\rho_k] &=& -\left[\Gamma (n_k-n_k^{(\mathrm{eq})}) \mathbb{I}+\frac{\Gamma}{2} (S_{k,\alpha}-S_{k,\alpha}^{(\mathrm{eq})})\sigma_\alpha\right. \nonumber \\
&&\left.+\frac{\Gamma_\perp}{2} S_{k,\beta}\sigma_\beta+\frac{\Gamma_\perp}{2} S_{k,\gamma}\sigma_\gamma\right], \label{Ic}
\end{eqnarray}
\begin{figure*}[htb]
  \includegraphics[width=18cm,angle=0]{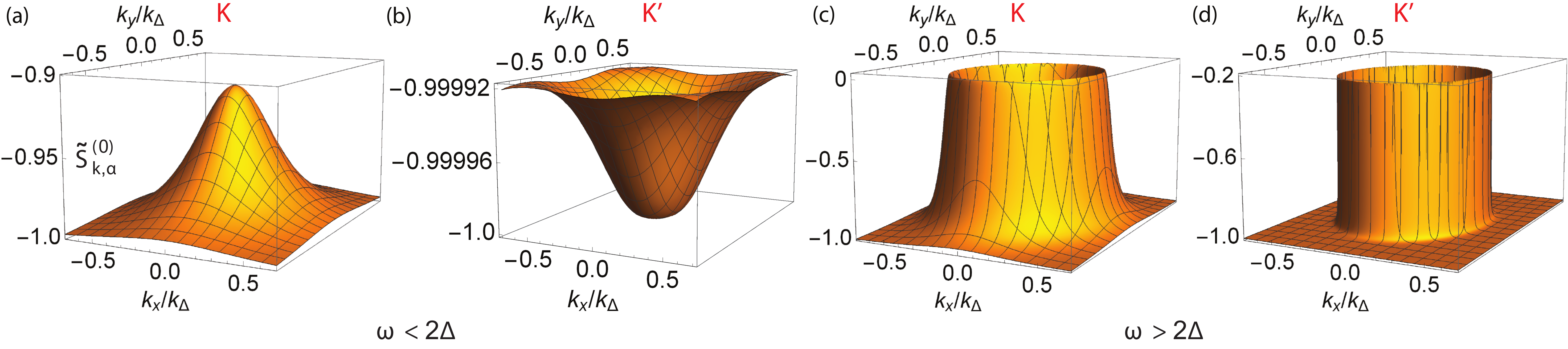}
  \caption{(Color online). Population difference $\tilde{S}_{k,\alpha}^{(0)} = S_{k,0}$ between the conduction and valence bands at valley K and K' under a circularly polarized pump field with helicity $\mu = 1$ and strength $E_0= 100$\; MV/m for (a)-(b) $\omega=1.4$ eV and (c)-(d) $\omega= 1.74$ eV. The labels for the K and K' valleys are indicated above the plots. Relaxation and dephasing parameters are taken as $\Gamma=\Gamma_\perp=1$ meV and the equilibrium band gap $2\Delta$ is the same as in Fig.~\ref{Ek1}.} 
\label{Sk014}
\end{figure*}
where $S_{k,\alpha}, S_{k,\beta}, S_{k,\gamma}$ denote the components of $\bm{S}_k$ along $\{\hat{\boldsymbol{\alpha}}_{k}, 
\hat{\boldsymbol{\beta}}_{k}, \hat{\boldsymbol{\gamma}}_{k}\}$, respectively. $S_{k,\alpha}$ describes the population difference $S_{k,\alpha} = \rho_{k,cc}-\rho_{k,bb}$  between the conduction band ($c$) and the valence band ($b$) and is also known as interband population inversion (with $S_{k,\alpha} = 1$ for full inversion), 
whereas $S_{k,\beta}, S_{k,\gamma}$ describe interband coherence that leads to optical polarization. $\Gamma$ and  $\Gamma_\perp$ phenomenologically capture the effects of 
the decay of interband population inversion and optical polarization as well as intraband momentum relaxation. We note that inclusion of dissipative effects are essential for the irradiated system to attain the non-equilibrium steady state. 
Before light is turned on, the system is assumed to be in equilibrium and the Fermi level is inside the band gap, with a completely filled valence band and an empty conduction band so that  $n_k^{(\mathrm{eq})} = 1/2$, $S_{k,\alpha}^{(\mathrm{eq})}=-1, S_{k,\beta}^{(\mathrm{eq})}=S_{k,\gamma}^{(\mathrm{eq})}=0$. 

\section{Effects of Pump Field: Zeroth-Order Density Matrix} \label{sec:p0th}

To obtain the photoconductivity, we solve Eq.~\eqref{QKEE} up to first order in $\bm{E}$ by linearizing the density matrix as $\rho_k=\rho_k^{(0)}+\rho_k^{(1)}$.  The density matrix $\rho_k^{(0)}$ is the zeroth-order  solution to Eq.~\eqref{QKEE} under a zero \textit{d.c.} probe field ${E} = 0$ and $\rho_k^{(1)}$ is the the first-order correction due to a finite ${E}$. Eq.~\eqref{QKEE} then reduces to the following two equations satisfied by $\rho_k^{(0)}$ and $\rho_k^{(1)}$:
\begin{eqnarray}  
\frac{\partial \rho_k^{(0)}}{\partial t}+i\left[H,\rho_k^{(0)}\right] &=& I_{\mathrm{s}}[\rho_k^{(0)}], \label{qkeq1} \\
\frac{\partial \rho_k^{(1)}}{\partial t}-e\bm{E}\cdot\frac{\partial \rho_k^{(0)}}{\partial \bm{k}}+i\left[H,\rho_k^{(1)}\right] &=& I_{\mathrm{s}}[\rho_k^{(1)}]. \label{qkeq2} 
\end{eqnarray}
Since we are interested in the steady-state regime, the above equations can be conveniently solved by transforming them into the rotating frame, in which the density matrix $\tilde{\rho}_k$ 
becomes time-independent within RWA: $\partial \tilde{\rho}_k/ \partial t = 0$. The resulting equations satisfied by $\tilde{\rho}_k^{(0)}$ and $\tilde{\rho}_k^{(1)}$ then take the same form as Eqs.~(\ref{qkeq1})-(\ref{qkeq2}) with $\partial/\partial t = 0$.

Our strategy for solving the $2\times 2$ kinetic equation Eq.~\eqref{qkeq1} in the pseudospin space is to project it onto the basis  $\left\{\mathbb{I},\sigma_\alpha,\sigma_\beta,\sigma_\gamma\right\}$,
which produces four linearly independent equations that can be solved simultaneously.  The zeroth and first order density matrices $\tilde{\rho}_k^{(0)}, \tilde{\rho}_k^{(1)}$ are then respectively expanded as
\begin{equation}
  \tilde{\rho}_k^{(0,1)}=n_k^{(0,1)}\mathbb{I}+\frac{1}{2}\left(\tilde{S}_{k,\alpha}^{(0,1)}\sigma_\alpha +\tilde{S}_{k,\beta}^{(0,1)}\sigma_\beta+\tilde{S}_{k,\gamma}^{(0,1)}\sigma_\gamma\right). \label{rho1}
\end{equation}
The rotating frame Hamiltonian $\tilde{H}$, written in the new pseudospin basis, has been derived in Eq.~\eqref{ham'}. Since the set of basis matrices satisfy the usual  commutation relation $[\sigma_i,\sigma_j]=2i\epsilon_{ijk}\sigma_k$ with $i,j,k \in \{\alpha, \beta,\gamma\}$, one can easily find   
\begin{eqnarray}
[\tilde{H}, \tilde{\rho}_k^{(0)}]&=&\left(\alpha_k-\frac{\omega}{2}\right)(\tilde{S}_\beta^{(0)}i\sigma_\gamma - \tilde{S}_\gamma^{(0)}i\sigma_\beta) \label{comm0th}\\
&&-\frac{\Lambda}{4}M_k\cos\phi (\tilde{S}_\alpha^{(0)}i\sigma_\gamma-\tilde{S}_\gamma^{(0)}i\sigma_\alpha) \nonumber \\
&&-\frac{\Lambda}{4}M_k\mu\sin\phi (\tilde{S}_\alpha^{(0)}i\sigma_\beta-\tilde{S}_\beta^{(0)}i\sigma_\alpha). \nonumber
\end{eqnarray}
Note that the charge density distribution function $n_k$ is decoupled from the kinetic equation for $\tilde{\bm{S}}_k$ since the contribution from $n_k$ vanishes in Eq.~\eqref{comm0th}  upon commutation operation. Substituting Eqs.~\eqref{rho1}-\eqref{comm0th} into the kinetic equation and solving, 
%Eq.~\eqref{qkeq1} and Eq.~\eqref{Ic} (with $\rho \to \tilde{\rho}$ and $\partial/\partial t = 0$ in the rotating frame) and solving the resulting equations,
we find the steady-state solution for 
$\tilde{\rho}_k^{(0)}$:
\begin{eqnarray}
\tilde{\rho}_k^{(0)} &=& n_k^{(\mathrm{eq})}\mathbb{I} + \frac{1}{2}S_{k,0}\sigma_\alpha+\frac{1}{2}(S_{k,1}\cos\phi+S_{k,2}\sin\phi)\sigma_\beta \nonumber \\
&&+\frac{1}{2}\mu(S_{k,2}\cos\phi-S_{k,1}\sin\phi)\sigma_\gamma, \label{rho0Sol}
\end{eqnarray}
where
\begin{eqnarray}
\left[\begin{array}{c}
        S_{k,0}\\
        S_{k,1} \\
        S_{k,2}\end{array}\right] 
  &=& \frac{-1}{\left(2\alpha_k - \omega\right)^2 + \Gamma_\perp^2 + {({\Lambda M_k}/{2})}^2 {\Gamma_\perp}/{\Gamma}} \nonumber \\
      &&\times\left[\begin{array}{c}
                                                                                                                                                             \left(2\alpha_k - \omega\right)^2 + \Gamma_\perp^2 \\
                                                                                                                                                             (\Lambda M_k/2)(2\alpha_k - \omega) \\
                                                                                                                                                             -(\Lambda M_k/2) \mu\Gamma_\perp \end{array}\right]. 
\label{Sk123}
\end{eqnarray}
Figs.~\ref{Sk014}(a)-(d) show the interband population difference $\tilde{S}_{k,\alpha}^{(0)} = S_{k,0}$ at valleys K and K' under a circularly polarized pump field with helicity $\mu = 1$ for the cases when the frequency is below and above the band gap. When $\omega < 2\Delta$ [Figs.~\ref{Sk014}(a)-(b)], a small population of electrons is excited into the conduction band localized around the band edge $k = 0$. Most of the electron population remains in the valence band, with $\tilde{S}_{k,\alpha}^{(0)} \approx -1$. For $\omega > 2\Delta$ [Figs.~\ref{Sk014}(c)-(d)], electrons of both valleys are excited predominantly
to those states that are peripheral to
the ring of resonant states $\omega = 2\alpha_k$ where the dynamical gap opens [Fig.~\ref{Ek1}(b)]. Near those states around the circular ``opening'' in Figs.~\ref{Sk014}(c) for valley K, $\tilde{S}_{k,\alpha}^{(0)}$ reaches a maximum of  $\sim -10^{-4}$ indicating that the valence band electrons there are strongly excited to the conduction band.
In comparison, less electrons are photoexcited at valley K' as shown in Figs.~\ref{Sk014}(d), where the maximum $\tilde{S}_{k,\alpha}^{(0)}$ reaches about $-0.3$. Because the dynamical gap is much smaller at K' than at K [Fig.~\ref{Ek1}(b)], the excited populations at K' are localized closely at the resonant states resulting in a much sharper distribution of $\tilde{S}_{k,\alpha}^{(0)}$ in the momentum space. 

\section{Effects of d.c. Bias: \\ First-order Density Matrix} \label{sec:p1st}

Having obtained the steady-state solution to Eq.\eqref{qkeq1}, we proceed to solve Eq.\eqref{qkeq2} in the rotating frame using the decomposition Eq.~(\ref{rho1}) for $\tilde{\rho}_k^{(1)}$. 
The \textit{d.c.} electric field is taken as $\bm{E}=E\hat{\bm{x}}$ directed along $\hat{\bm{x}}$. The $\bm{E}$-dependent driving term in Eq.~\eqref{qkeq2}  is completely determined by $\tilde{\rho}_k^{(0)}$ and can be resolved as 
\begin{eqnarray} \label{Ddef}
e\bm{E}\cdot\frac{\partial \tilde{\rho}_k^{(0)}}{\partial \bm{k}}
&=&eE\bigg(\mathcal{D}_\mathbb{I}\mathbb{I}+\mathcal{D}_{k,\alpha}\sigma_\alpha+\mathcal{D}_{k,\beta}\sigma_\beta+\mathcal{D}_{k,\gamma}\sigma_\gamma\bigg), \nonumber \\
\end{eqnarray}
with functions $\mathcal{D}_{\mathbb{I}}, \mathcal{D}_{k,\alpha}, \mathcal{D}_{k,\beta}, \mathcal{D}_{k,\gamma}$ as coeffcients. From Eqs.~\eqref{rho0Sol}-\eqref{Sk123} it is obvious that $\mathcal{D}_{\mathbb{I}} = 0$, and we can obtain explicit expressions of $\mathcal{D}_{k,\alpha}, \mathcal{D}_{k,\beta}, \mathcal{D}_{k,\gamma}$ as provided in Appendix A.  
The commutator $[\tilde{H}, \tilde{\rho}_k^{(1)}]$ is the same as in Eq.\eqref{comm0th} with the superscript $(0)$ replaced by $(1)$. 
It follows that $n_k^{(1)} = 0$ and 
$\tilde{S}_{k,\alpha}^{(1)},\tilde{S}_{k,\beta}^{(1)},\tilde{S}_{k,\gamma}^{(1)}$ are determined by
\begin{eqnarray}
&&\begin{bmatrix}
\Gamma & -\frac{\Lambda}{2}M_k\mu\sin\phi & -\frac{\Lambda}{2}M_k\cos\phi \\
\frac{\Lambda}{2}M_k\mu\sin\phi &\Gamma_\perp& \left(2\alpha_k -\omega\right) \\
\frac{\Lambda}{2}M_k\cos\phi &- \left(2\alpha_k -\omega\right)&\Gamma_\perp 
\end{bmatrix}
\begin{bmatrix}
\tilde{S}_{k,\alpha}^{(1)} \\ \tilde{S}_{k,\beta}^{(1)}\\ \tilde{S}_{k,\gamma}^{(1)}
\end{bmatrix} \nonumber \\
                 &=&2eE\begin{bmatrix} \mathcal{D}_{k,\alpha} \\ \mathcal{D}_{k,\beta} \\ \mathcal{D}_{k,\gamma} \end{bmatrix}. \label{matrixBCD}
\end{eqnarray}
\begin{figure*}[htb]
  \includegraphics[width=18cm,angle=0]{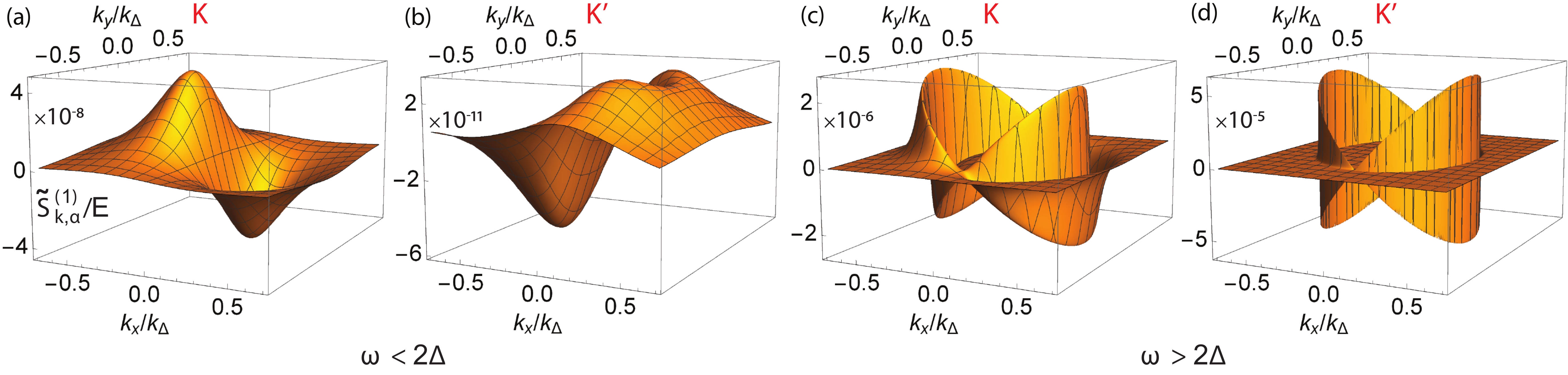}
  \caption{(Color online). First-order correction to the population difference $\tilde{S}_{k,\alpha}^{(1)}/E$ scaled by the \textit{d.c.} bias field $E$ between the conduction and valence bands at valley K and K' for (a)-(b) $\omega=1.4$ eV and (c)-(d) $\omega= 1.74$ eV. The labels for the K and K' valleys are indicated above the plots. The pump field has the same helicity and strength, and the values of $\Gamma,\Gamma_\perp,\Delta$ are the same as in Fig.~\ref{Sk014}.}
\label{Sk015}
\end{figure*}

The above equation gives explicit analytic expressions for $\tilde{S}_{k,\alpha}^{(1)}, \tilde{S}_{k,\beta}^{(1)}, \tilde{S}_{k,\gamma}^{(1)}$, which are relegated in Appendix B.
In Figs.~\ref{Sk015}(a)-(d), we show the correction to the population difference $\tilde{S}_{k,\alpha}^{(1)}$ due to the \textit{d.c.} electric field at both valleys for the $\omega$ below and above the band gap. Since $\tilde{S}_{k,\alpha}^{(1)}$ is proportional to $E$, we plot $\tilde{S}_{k,\alpha}^{(1)}/E$. In contrast to $\tilde{S}_{k,\alpha}^{(0)}$, $\tilde{S}_{k,\alpha}^{(1)}$ is asymmetric in $k$-space due to  the \textit{d.c.} field breaking in-plane rotational symmetry. Below the band gap [Fig.~\ref{Sk015}(a)-(b)], $\tilde{S}_{k,\alpha}^{(1)}$ is generally very small. For a \textit{d.c.} field $E = 10\,\mathrm{kV/m}$ for instance, $\tilde{S}_{k,\alpha}^{(1)} \sim 10^{-4}$ at valley K and $\tilde{S}_{k,\alpha}^{(1)} \sim 10^{-7}$ at valley K'. When the frequency is increased to above the band gap, $\tilde{S}_{k,\alpha}^{(1)}$ is dramatically enhanced near the resonant states by two and six orders of magnitude respectively as seen in Fig.~\ref{Sk015}(c)-(d). 
This shows that a resonant pump field excitation induces a much stronger effect on the photoexcited population distribution perturbed by the \textit{d.c.} bias. 
%more prominent response on the 
%the interplay between the strong \textit{a.c} field and the weak \textit{d.c.} bias induces a more prominent response on the photoexcited population for resonant pump field excitations. 
%near the resonant transition.
%the pump field has a much stronger effect on the \textit{d.c.} field-correction to the photoexcited population when the frequency is above the band gap.

The degree of asymmetry can be analyzed by resolving $\tilde{S}_{k,\alpha}^{(1)}$ into even and odd harmonics of $\phi$. 
%$\sin\phi$ and $\cos\phi$.
While Figs.~\ref{Sk015}(a)-(d) seem to show only an asymmetry along the $k_x$  direction, there is also a small degree of asymmetry along the $k_y$ direction that is not apparent at the scale of the plots. In Appendix B we show the explicit expressions of the first odd ($\sin\phi$) and even ($\cos\phi$) harmonics of $\tilde{S}_{k,\alpha}^{(1)}$, which corresponds to asymmetries along the $k_y$ and $k_x$ directions respectively. 
As we will explain in Sec.~\ref{sec:dcI}, 
the asymmetry of this distribution function along the transverse direction to the \textit{d.c.} bias, along with smaller effects from the interband coherences $\tilde{S}_{k,\beta}^{(1)}$ and  $\tilde{S}_{k,\gamma}^{(1)}$, leads to the photo-induced anomalous Hall effect.

The preferential coupling between the left (right) circularly polarized light and the K (K') valley results in a population imbalance of photoexcited connduction band electrons between the two valleys.
Using the pseudospin-to-band unitary transformation $\mathcal{U}_k$, the conduction band density matrix can be found as $\rho_{k,cc} = n_k+\tilde{S}_{k,\alpha}/2$. $\rho_{k,cc}$ is predominantly given by the zeroth order contribution $n_k^{\rm (eq)}+{S}_{k,0}/2$ as the correction $\tilde{S}_{k,\alpha}^{(1)}/2$ induced by the \textit{d.c.} bias is comparatively small. Because $n_k^{\rm (eq)}=1/2$ is independent of the valley degrees of freedom, the conduction band population difference between the two valleys is $\Delta n_{\rm v}=
\sum_k(\rho_{k,cc}^{K}-\rho_{k,cc}^{K'})  =\sum_k[(S_{k,0}+\tilde{S}_{k,\alpha}^{(1)})^{K}-(S_{k,0}+\tilde{S}_{k,\alpha}^{(1)})^{K'} ]/2$. 
\begin{figure}[htb] %[hp!t!b]
  \includegraphics[width=8.5cm,angle=0]{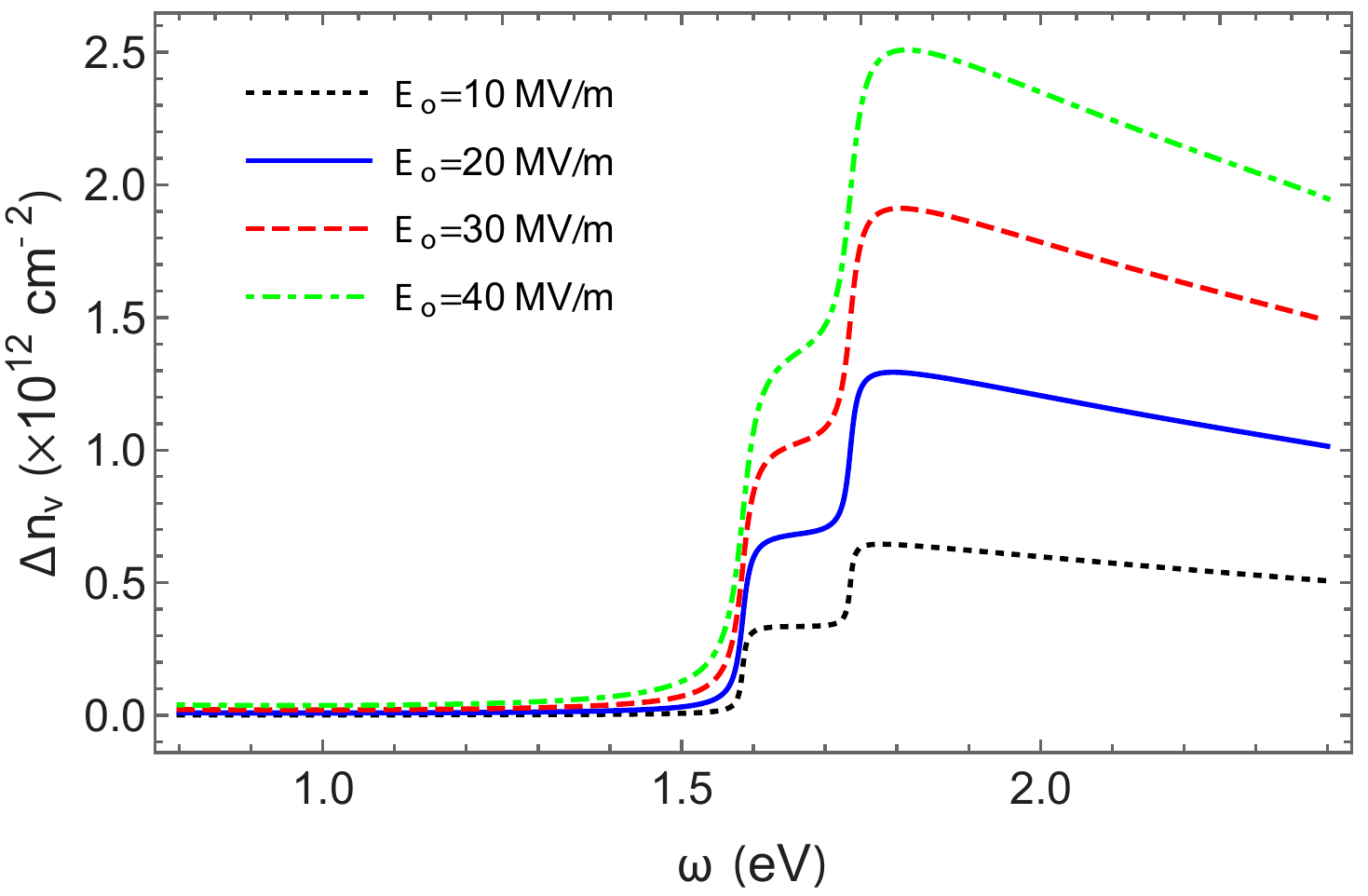}
  \caption{(Color online). Conduction band population difference between the two valleys to the zeroth order, as a function of $\omega$ for different values of $E_0$ with  $\mu=1$.
    Relaxation and dephasing parameters are taken as $\Gamma=\Gamma_\perp=1$ meV.}
\label{deltanv}
\end{figure}
Then the total population imbalance can be found by summing over the spin degrees of freedom in the original TMD Hamiltonian Eq.~\ref{h0}, which correspond to the two values of the gap $2\Delta_1$ and $2\Delta_2$. They give the interband transition energies at $k = 0$ for the two spins. Fig.~\ref{deltanv} shows the resulting total $\Delta n_{\rm v}$ as a function of the frequency for different values of the pump field. $\Delta n_{\rm v}$ exhibits a shoulder-like feature when the frequency reaches $2\Delta_1$ and then peaks at the second gap $2\Delta_2$, before tailing off gradually at higher frequencies. At this point, it may be tempting to obtain the anomalous Hall conductivity from this valley population imbalance as in the \textit{d.c.} case. However, because of the formation of photon-dressed bands in the presence of a pump field, the photo-induced Hall current is no longer simply given by this valley population imbalance and the Berry curvatures of the equilibrium bands. We can estimate the Hall conductivity obtained in this way \cite{mak2014valley} using Fig.~\ref{deltanv} and find that it is an order of magnitude too small compared to our exact results presented in Fig.~\ref{Jyt1t2}. 
Instead, the photo-induced transport currents are determined by the distribution function $\tilde{\bm{S}}_{k}^{(1)}$ of the photon-dressed bands as described below.

\section{Longitudinal and Anomalous Hall Photoconductivities} \label{sec:dcI}

\begin{figure*}[htb]
  \includegraphics[width=18cm,angle=0]{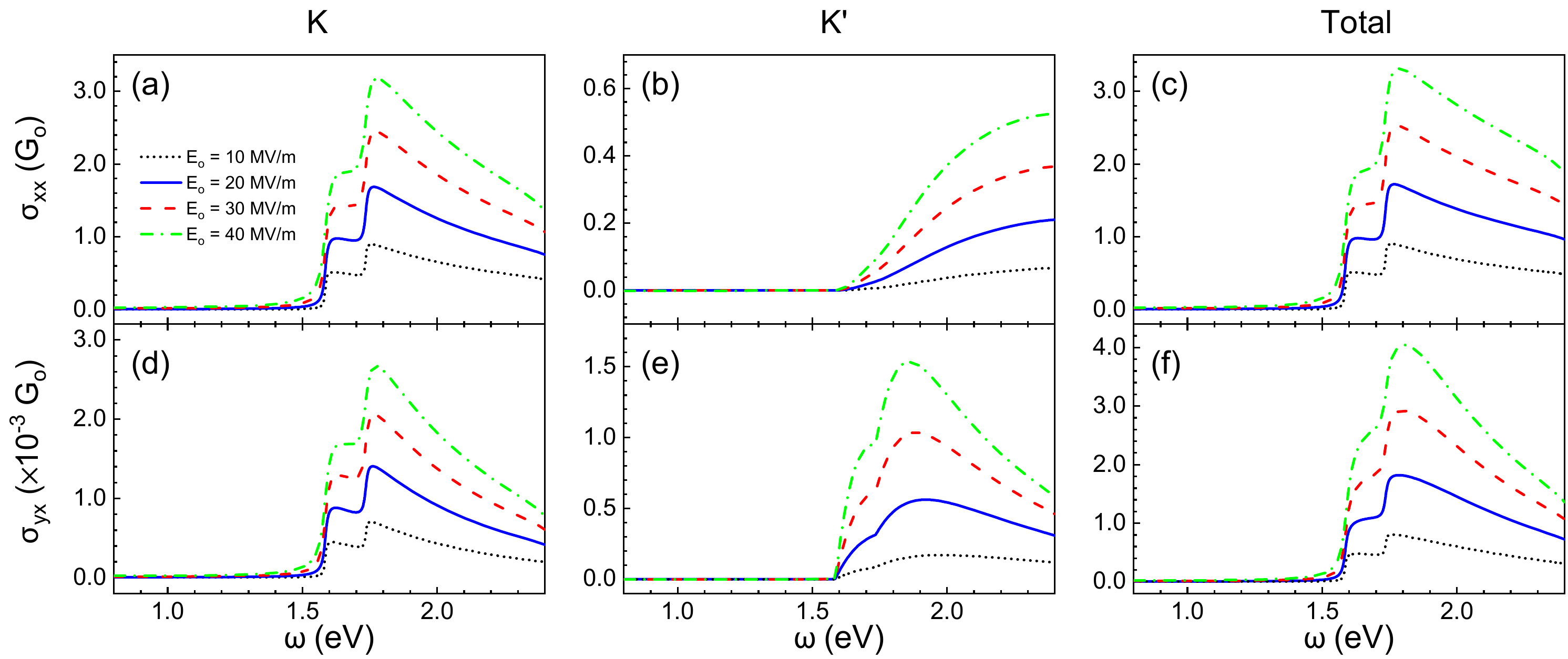}
  \caption{(Color online). Longitudinal $\sigma_{xx}$ and Hall $\sigma_{yx}$ conductivities in units of $G_0=e^2/\hbar$ as a function of $\omega$ under different strengths of pump field $E_0$ with helicity $\mu=1$. Panels  (a) and (d) show the contributions due to the K valley ($\tau =1$) while (b) and (e) show the K' valley ($\tau =-1$), and panels (c) and (f) show the sum of the two valleys' contributions. Relaxation and dephasing rates are the same as in Fig.~\ref{deltanv}. } \label{Jyt1t2}
\end{figure*}
To calculate the photovoltaic current, the density matrix needs to be transformed back into the stationary frame $\rho_k^{(1)}=U\tilde{\rho}_k^{(1)}U^{\dagger}$, 
\begin{eqnarray}
\rho_k^{(1)} &=& n_k^{(1)}\mathbb{I} + \frac{1}{2}\tilde{S}_{k,\alpha}^{(1)}\sigma_\alpha + \frac{1}{2}\left( \tilde{S}_{k,\beta}^{(1)} \cos\omega t- \tilde{S}_{k,\gamma}^{(1)}\sin\omega t\right)\sigma_\beta \nonumber \\
&&+\frac{1}{2}\left(\tilde{S}_{k,\beta}^{(1)}\sin\omega t  + \tilde{S}_{k,\gamma}^{(1)}\cos\omega t \right)\sigma_\gamma. \label{lf1st}
\end{eqnarray}
The expectation value of the current density is then calculated from $\bm{J}=\Sigma_k \mathrm{Tr}\{\bm{j}(t) \rho_k^{(1)}(t)\}$, where `Tr' denotes trace over degrees of freedom other than the  momentum, and $\bm{j}(t)$ is the single-electron current operator, 
\begin{eqnarray} \label{jkop}
\bm{j}(t)=-e\frac{\partial H_{\mathrm{R}}(t)}{\partial \bm{k}}=-e\left(\frac{\partial H_{\mathrm{R}}}{\partial k}\hat{\bm{k}}+\frac{1}{k}\frac{\partial H_{\mathrm{R}}}{\partial \phi}\hat{\bm{\phi}}\right).
\end{eqnarray}
$H_{\mathrm{R}}(t)$ above is the stationary-frame Hamiltonian \textit{within the RWA}, and can be obtained by transforming $\tilde{H}$ in Sec. \ref{sec:Ham} back to the stationary frame $H_{\mathrm{R}}(t) = U\tilde{H}U^{\dagger}-iU\partial_t U^{\dagger}$,
\begin{eqnarray}
H_{\mathrm{R}}(t) 
&=&\alpha_k\sigma_\alpha+\frac{\Lambda}{4}M_k\left(\cos\phi\cos\omega t+\mu\sin\phi\sin\omega t\right)\sigma_\beta \nonumber \\
&& + \frac{\Lambda}{4}M_k\left(\cos\phi\sin\omega t-\mu\sin\phi\cos\omega t\right)\sigma_\gamma. \label{ham'_stat}
\end{eqnarray}

The matrix trace calculation can be facilitated by decomposing the  longitudinal ($x$-direction) and transverse ($y$-direction) single-electron current operators into components of $\{\sigma_\alpha, \sigma_\beta, \sigma_\gamma\}$, such that $j_{i}(t)=\hat{\bm{i}}\cdot\bm{j}_k(t)=j_{i,\alpha}(t)\sigma_\alpha+j_{i,\beta}(t)\sigma_\beta+j_{i,\gamma}(t)\sigma_\gamma$ with  $i \in \{x,y\}$. Explicit expressions of $j_{i,\alpha}(t), j_{i,\beta}(t), j_{i,\gamma}(t)$ are relegated to Appendix A. It is easy to verify that the basis matrices satisfy the trace relation $\text{Tr}\left\{\sigma_{\mu}\sigma_{\nu}\right\}=2\delta_{\mu\nu}$, where $\mu,\nu \in \{\alpha,\beta,\gamma\}$. Using this property with Eq.~\eqref{lf1st}, the photovoltaic longitudinal and Hall currents can be calculated from $\rho_k^{(1)}$ as 
\begin{eqnarray}
J_{i} &=&\sum_k \bigg[\tilde{S}_{k,\alpha}^{(1)}j_{i,\alpha}+ \left(\tilde{S}_{k,\beta}^{(1)}\cos\omega t  -\tilde{S}_{k,\gamma}^{(1)}\sin\omega t\right )j_{i,\beta} \nonumber \\
&&\quad+\left(\tilde{S}_{k,\beta}^{(1)}\sin\omega t  +\tilde{S}_{k,\gamma}^{(1)} \cos\omega t\right)j_{i,\gamma}\bigg].  \label{Jitau}  
\end{eqnarray}

Before proceeding to calculate the photoconductivities, it is useful to first check that our formulation recovers the correct dark conductivity. The scenario of vanishing pump field corresponds to taking the limit $\Lambda,\omega \to 0$ such that $\Lambda/\omega \to 0$. The rotating frame reduces to the stationary frame and the Hamiltonian in Eq.~\eqref{ham'} becomes the original Hamiltonian without light  $H=\alpha_k\sigma_\alpha$. Damping terms $\Gamma, \Gamma_{\perp} $ can be taken as zero because the Fermi energy lies within the band gap. Solutions to Eq.~\eqref{matrixBCD} then reduce to   
\begin{eqnarray}
\tilde{S}_{k,\alpha}^{(1)}&=&0, \\
\tilde{S}_{k,\beta}^{(1)}&=&-\frac{\tau eE}{2k\alpha_k}\sin\theta_k\cos\theta_k\cos\phi, \\
\tilde{S}_{k,\gamma}^{(1)}&=&\frac{eE}{2k\alpha_k}\sin\theta_k\sin\phi.
\end{eqnarray}
From Eq.~\eqref{lf1st}, the first-order density matrix then becomes 
\begin{eqnarray}
\rho_k^{(1)}&=&n_k^{(1)}\mathbb{I}-\frac{\tau eE}{4k\alpha_k}\sin\theta_k\left(\cos\theta_k\cos\phi \sigma_\beta-\tau\sin\phi\sigma_\gamma\right). \nonumber \\
\label{rhok1}
\end{eqnarray}
The $y$-component of the single-electron current operator in Eq.~\eqref{jkop} reduces to $-ev\sigma_y$, which when written in pseudospin basis is  
\begin{eqnarray}
j_y&=&-e\frac{\alpha_k\sin\theta_k}{k}\left(\sin\theta_k\sin\phi \sigma_\alpha+\cos\phi\sigma_\beta \right. \nonumber \\
&&\left.-\tau\cos\theta_k\sin\phi \sigma_\gamma\right). \label{jy0}
\end{eqnarray}
Calculating the transverse current $J_y=\Sigma_k \mathrm{Tr}\{j_y \rho_k^{(1)}\}$, we recover the well-known dark valley-resolved Hall conductivity $\sigma_{yx}^{\tau} = {J_y}/{E} = {\tau e^2}/{4\pi\hbar}$ 
where the superscript $\tau$ distinguishes the contribution from each valley. 
%and we have restored $\hbar$.
Similarly, we find a vanishing longitudinal conductivity $\sigma_{xx}^\tau = 0$ for a vanishing pump field, as expected for undoped TMDs.

\begin{figure}[htb]
  \includegraphics[width=8.5cm,angle=0]{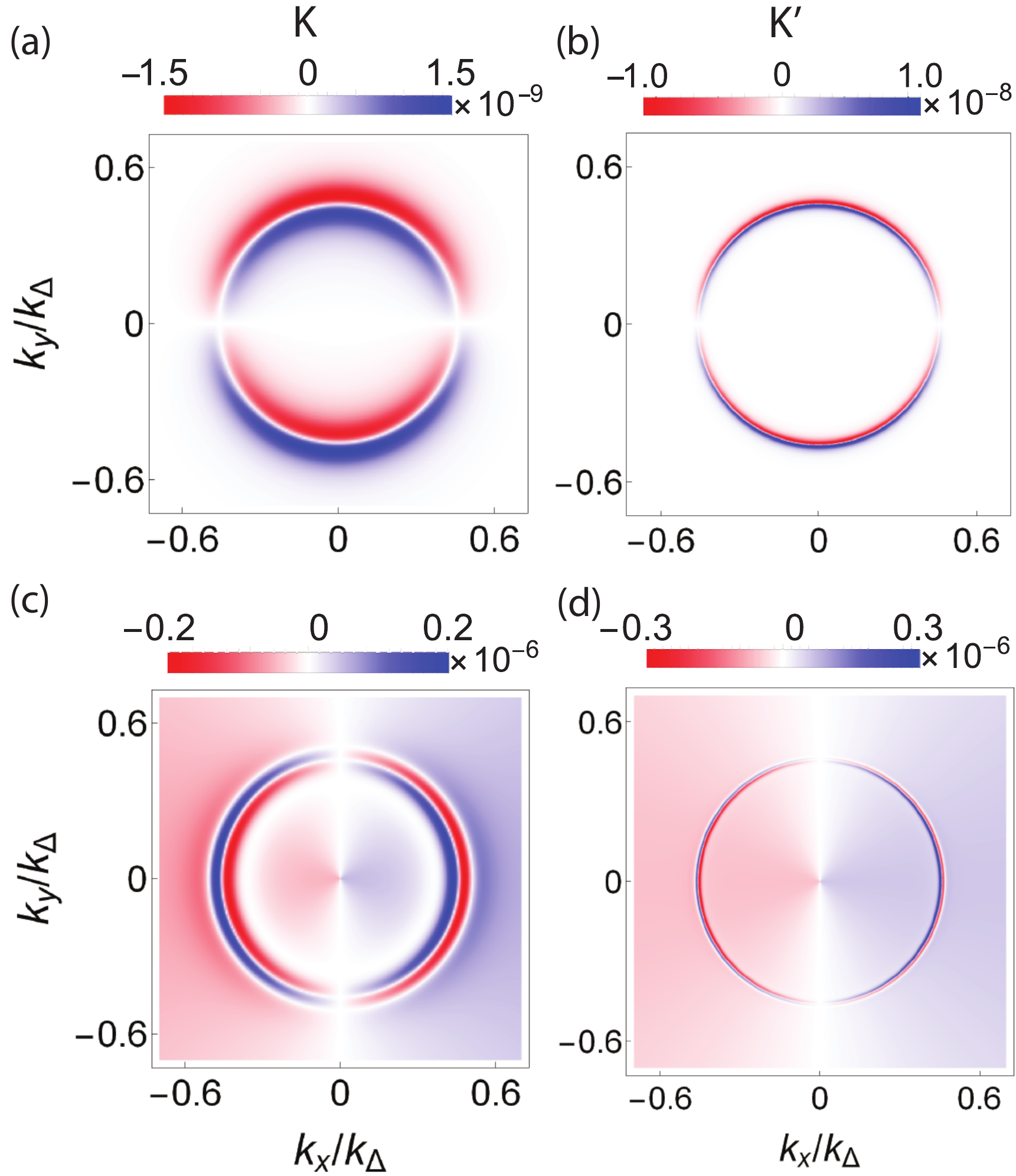}
  \caption{(Color online). First harmonic components of $\tilde{S}_{k,\alpha}^{(1)}/E$ at $E_0 = 100\,\mathrm{MV/m}$, $\omega = 1.74\,\mathrm{eV}$ and  helicity $\mu=1$. The $\sin\phi$ component is shown in (a) for valley K and (b) for valley K', while the the $\cos\phi$ component is shown in (c) for valley K and (d) for valley K'. Relaxation and dephasing rates are taken as $\Gamma = \Gamma_{\perp} = 7\,\mathrm{meV}$. } \label{sincos}
\end{figure}
We now return to Eq.~(\ref{Jitau}). Subtracting off the dark current contribution and integrating over one time period,  
we obtain the following expressions for the time-averaged 
photo-induced longitudinal and Hall currents for spin $s$ and valley $\tau$:
\begin{eqnarray} 
 J_{x} 
&=&\sum_k\frac{\sin^2\theta_k}{k}\left\{\tilde{S}_{k,\alpha}^{(1)}\alpha_k\cos\phi+\frac{\mu\Lambda}{8}\tilde{S}_{k,\gamma}^{(1)}M_{k,+}\sin2\phi \right.\label{JxtauM} \nonumber \\
&&\left.\quad+\frac{\Lambda}{8}\tilde{S}_{k,\beta}^{(1)} \left[M_{k,-}-M_{k,+}\cos2\phi\right]\right\}, 
\end{eqnarray}
\begin{eqnarray}
 J_{y} &=&\sum_k\frac{\sin^2\theta_k}{k}\left\{\tilde{S}_{k,\alpha}^{(1)}\alpha_k\sin\phi-\frac{\Lambda}{8}\tilde{S}_{k,\beta}^{(1)} M_{k,+}\sin2\phi\right. \label{JytauM} \nonumber \\
&&\left.\quad-\frac{\mu\Lambda}{8}\tilde{S}_{k,\gamma}^{(1)}\left[M_{k,-}+M_{k,+}\cos2\phi\right]\right\},
\end{eqnarray}
where $M_{k,\pm}= \mu\tau\pm\cos\theta$. In Eqs.~\eqref{JxtauM}-\eqref{JytauM} above, the first term dependent on $\tilde{S}_{k,\alpha}^{(1)}$ corresponds to a Drude-like intraband response from the  photon-dressed conduction and valance bands, whereas the second and third terms dependent on $\tilde{S}_{k,\beta}^{(1)}, \tilde{S}_{k,\gamma}^{(1)}$ arise from interband coherence effects. Because of the momentum integration, it is clear that only the first odd (even) harmonic of $\tilde{S}_{k,\alpha}^{(1)}$ contributes to the intraband response of $J_y$ ($J_x$), while only the zeroth, second odd and even harmonics of $\tilde{S}_{k,\beta}^{(1)}, \tilde{S}_{k,\gamma}^{(1)}$ enter into the interband coherence contributions of $J_y$ and $J_x$. The total longitudinal and Hall photoconductivities are finally obtained by summing Eqs.~\eqref{JxtauM}-\eqref{JytauM} over the spin and valley degrees of freedom and dividing over the \textit{d.c.} probe field $E$. In  the \textit{d.c.} anomalous Hall effect, interband coherences give rise to the intrinsic geometric contribution in ferromagnetic metals and in particular to quantized topological contribution in magnetic insulators \cite{AHE_RMP,AHE_Rev2}. 
%currents are finally obtained by summing Eqs.~\eqref{JxtauM}-\eqref{JytauM} over the spin and valley degrees of freedom, and by dividing over the \textit{d.c.} probe field $E$ they give the corresponding longitudinal and Hall conductivities. 

Fig.~\ref{Jyt1t2} shows our numerical results for  the valley-specific and total photoconductivities under left circularly polarized light $(\mu=1)$ calculated from Eqs.~(\ref{JxtauM})-(\ref{JytauM}). One first notices that the K valley contribution is larger than that of the K' valley for both the longitudinal [Figs.~\ref{Jyt1t2}(a)-(b)] and Hall conductivities [Figs.~\ref{Jyt1t2}(d)-(e)]. Similar to $\Delta n_{\mathrm{v}}$, the shoulder and peak features at $\omega = 2\Delta_1$ and $2\Delta_2$ are clearly visible for $\sigma_{xx}$ and $\sigma_{yx}$ at valley K, while they are less prominent for the conductivities at valley K'. Interestingly, we find that the photo-induced  $\sigma_{yx}$ at the two valleys carry the same sign, in contrast to the unpumped case where different valley contributions to the dark Hall conductivity have opposite signs. The underlying reason can be seen as follows. 
% which we can understand as follows.

In Eq.~\eqref{JytauM} for  the Hall conductivitiy, the contributions from interband coherences $\tilde{S}_{k,\beta}^{(1)}, \tilde{S}_{k,\gamma}^{(1)}$ are typically small compared to the contribution due to  population inversion  $\tilde{S}_{k,\alpha}^{(1)}$, as shown in Appendix C. Moreover, these interband coherence terms are dominated by their K valley contributions, which are larger than the  corresponding K' contributions by two orders of magnitude. Therefore, the valley dependence of $\sigma_{yx}$ is principally due to the intraband response term from $\tilde{S}_{k,\alpha}^{(1)}$. 
%the effect of different valleys on $\sigma_{yx}$ is mainly dependent on the population inversion term $\tilde{S}_{k,\alpha}^{(1)}$.
Figs.~\ref{sincos}(a)-(b) show an intensity plot of the first odd harmonic component of $\tilde{S}_{k,\alpha}^{(1)}$ that contributes to the Hall conductivity through Eq.~\ref{JytauM}. One can see that $\tilde{S}_{k,\alpha}^{(1)}$ at valleys K and K' [panels (a) and (b)] share the same sign as indicated by the same color at every $k$-point, and thus contribute  to the photo-induced Hall current with the same sign. 
In the case of the  longitudinal conductivity in Eq.~\eqref{JxtauM}, we find that  the intraband contribution dominates over the contributions from interband coherences
%$\tilde{S}_{k,\beta}^{(1)}, \tilde{S}_{k,\gamma}^{(1)}$,
so $\sigma_{xx}$   is largely contributed by the $\cos\phi$ harmonic component of $\tilde{S}_{k,\alpha}^{(1)}$. 
As shown in  Fig.~\ref{sincos}(c)-(d), the first even harmonic also  shares the same sign between the two valleys but is  generally much larger than the first odd harmonic.

Returning to Fig. \ref{Jyt1t2}, panels (c) and (f) show the total conductivities obtained from summing the two valleys' contributions. The magnitude of $\sigma_{xx}$ is about three orders of magnitude larger than that of $\sigma_{yx}$. If the circular polarization state is changed from $\mu = 1$ to $\mu = -1$, our numerical results show that both the magnitude and sign of the longitudinal conductivity remain unchanged, while the valley-specific contributions of the Hall conductivity are changed according to $\sigma_{yx}^{\tau=\mp1,\mu=-1}=-\sigma_{yx}^{\tau=\pm 1,\mu=1}$, resulting in an overall sign change of the total Hall conductivity $\sigma_{yx}$ as expected on  grounds of time-reversal.

To summarize, we find that both the photo-induced anomalous Hall and longitudinal conductivities are chiefly due to the intraband response of photon-dressed electrons arising from their asymmetric momentum-space distribution functions, accompanied by generally smaller contributions due to interband coherences. The latter, which correspond to the off-diagonal elements of the density matrix in the band representation, are the origin of geometric effects and give rise to Berry curvatures \cite{Berry1,Berry2,Berry3}. Hence our findings imply that the intrinsic geometric contribution only plays a secondary role in photo-induced anomalous Hall effect, in contrast to the case of \textit{d.c.} valley Hall effect \cite{valley_Niu_1,xiao2012coupled}. Our findings here are consistent with Ref.~\cite{AHE_Exp8} that have reached a similar conclusion in photoexcited graphene.  

%We note that asymmetry in the population distribution is also reported to be the main contribution in the photo-induced anomalous Hall effect in graphene \cite{AHE_Exp8}, while the Berry curvature contribution   computed from the instantaneous eigenstates was found to be small.
%The fact that interband coherence effects can also be partly captured by Berry curvatures \cite{Berry1,Berry2} suggests that our findings here and that of Ref.~\cite{AHE_Exp8} are consistent.  
%in qualitative agreement. 
%This is consistent with our findings here, since interband coherence effects can also be partly captured by Berry curvatures in an effective theory within a truncated band Hilbert space \cite{Berry1,Berry2}. 

In this work we have provided a non-interacting theory for the photo-induced anomalous Hall effect, neglecting the effects of excitons and trions. This is justified for the reason that 
excitons under a \textit{d.c.} bias are rapidly dissociated into free electrons and holes \cite{Exciton_1,ubrig2017microscopic} that contribute to steady-state transport. 
Trion effects, on the other hand, do not contribute in undoped samples we are considering where the equilibrium Fermi level lies deep within the band gap. Excitonic effects, however, could contribute in a more subtle way. In systems whose low-energy Hamiltonian breaks Galilean invariance, excitonic effects couple the intraband and interband dynamics resulting in  interaction-induced correction in dynamic transport properties such as the Drude weight \cite{Excitonic1,Excitonic3}. This effect is strongest in gapless systems such as graphene and is generally suppressed with increasing band gap \cite{Excitonic2}. Although TMDs have a large band gap, their electron-electron interaction effect is also stronger than in graphene or gapped bilayer graphene, and further study could shed light on whether the competition between these two effects would lead to considerable interaction correction to the anomalous Hall conductivity. In this paper we have considered only the intrinsic band  structure  contribution to the photo-induced anomalous Hall effect. A further extension of our theory could include the extrinsic effect due to spin-orbit scattering with impurities \cite{AHE_RMP}, which will be a subject of future investigation. 
%Also, we have only considered the intrinsic anomalous Hall effect in this work. A further extension of our work could include the extrinsic effect due to spin-orbit scattering with impurities.
Finally, we emphasize that while we are motivated by TMDs in this work, the theoretical method we  developed for the massive Dirac model and its massless limit 
can be applied more generally to other materials with gapped or gapless Dirac quasiparticles  \cite{DM_Rev,WSM_Rev,TI_Rev}
driven by a strong pump field. 

\section{Conclusion} \label{sec:concl}

To close, we have presented a theory for the photo-induced valley Hall transport for undoped 2D transition-metal dichalcogenides under a strong optical pump field. Our theory is developed using the density matrix formalism that enables treatment of the photon-dressed bands and carrier kinetics on an equal footing. The conceptual simplicity of our method allows to obtain useful theoretical insights on the population distribution of the photon dressed bands. Under circularly polarized pump field, we find considerable differences in the photon-dressed bands and the non-equilibrium carrier distributions at the two  valleys due to the valley-dependent optical selection rule. In each valley, electrons are predominantly excited to photon-dressed states  around the dynamical gap. Both the valley polarization and the photo-induced anomalous Hall conductivity are found to increase with the pump field and display notable signatures at the spin-resolved interband (\textit{i.e.} `A' and `B') transition energies. Despite this similiarity, we show that valley polarization plays a less important role in causing photo-induced Hall effect than was commonly assumed,
%in their behaviors as a function of field and frequency,
%our theory elucidate that valley polarization plays a less important role in photo-induced Hall effect  than was commonly assumed,
and the Hall effect is mainly driven by an asymmetric momentum-space distribution of photon-dressed electrons in the transverse direction. The theory and findings presented in this work highlight the important role of photon-dressed bands in understanding photo-induced transport, and demonstrate the viability of optical control of spins and valleys through the photon dressing effects of electronic bands.
%anomalous Hall transport through the photon dressing effects of electronic bands.

\begin{acknowledgments}
We thank Ben Yu-Kuang Hu and Patrick Kung for useful discussions. This work was supported by the U.S. Department of Energy, Office of Science, Basic Energy Sciences under Early Career Award No.  DE-SC0019326 and by the Research Grants Committee funds from the University of Alabama. 
\end{acknowledgments}

\begin{widetext}

\section{Appendix} \label{sec:append}
  
\subsection{Driving term $\mathcal{D}$ and single-particle current operators $j_x$ and $j_y$} \label{app:Djcomp}
In this appendix we provide explicit analytic expressions for the quantities too lengthy to be included in the main text. By decomposing the driving term in the kinetic  equation as in Eq. \eqref{Ddef} into the identity and transformed Pauli matrices, we have  
\begin{eqnarray}
\mathcal{D}_{k,\alpha}&=&\frac{1}{2}\left\{\cos\phi\left[\frac{\partial S_{k0}}{\partial k}+\mu\tau\frac{\partial \theta_k}{\partial k}(S_{k2}\cos\phi-S_{k1}\sin\phi)\right]+\frac{1}{k}\sin\phi\sin\theta(S_{k1}\cos\phi+S_{k2}\sin\phi)\right\}, \\
%%%
\mathcal{D}_{k,\beta}&=&\frac{1}{2}\left\{\cos\phi\left(\frac{\partial S_{k1}}{\partial k}\cos\phi+\frac{\partial S_{k2}}{\partial k}\sin\phi\right)-\frac{1}{k}\sin\phi\left[\sin\theta S_{k0}-\mu\tau(S_{k1}\sin\phi-S_{k2}\cos\phi)M_{k,-}\right]\right\}, \\
%%%
\mathcal{D}_{k,\gamma}&=&\frac{1}{2}\left\{\cos\phi\left[\mu\left(\frac{\partial S_{k2}}{\partial k}\cos\phi-\frac{\partial S_{k1}}{\partial k}\sin\phi\right)-S_{k0}\tau\frac{\partial \theta_k}{\partial k}\right]+\frac{1}{k}\tau\sin\phi\left(S_{k1}\cos\phi+S_{k2}\sin\phi\right)M_{k,-}\right\}.
\end{eqnarray}
%%%%%%%%
The single-particle current operator is calculated in the stationary frame from the Hamiltonian in Eq. \eqref{ham'_stat} obtained within the RWA, 
%%%%
\begin{eqnarray}
j_{x,\alpha}&=&-e\left[\frac{\partial \alpha_k}{\partial k} \cos\phi+\frac{\Lambda}{8}(M_{k,+})^2 \frac{1}{k}\tau\sin\theta\sin\omega t+\frac{\Lambda}{8}\frac{1}{k}\tau\sin^3\theta(-\sin\omega t \cos 2\phi+\mu\sin 2 \phi\cos\omega t)\right], \\
j_{x,\beta}&=&-e\left[-\alpha_k\sin\theta\frac{1}{k}\sin\phi+\frac{\Lambda}{8}\frac{1}{k}\sin^2\theta M_{k,-}\cos\omega t-\frac{\Lambda}{8}\frac{1}{k}\sin^2\theta M_{k,+}(\cos\omega t \cos 2 \phi+\mu\sin 2 \phi\sin\omega t)\right], \\
j_{x,\gamma}&=&-e\left[-\tau\alpha_k\sin\theta\cos\theta\frac{1}{k}\cos\phi+\frac{\Lambda}{8}\frac{1}{k}\sin^2\theta M_{k,-}\sin\omega t+\frac{\Lambda}{8}\frac{1}{k}\sin^2\theta M_{k,+}(\mu\cos\omega t \sin 2 \phi-\cos 2 \phi\sin\omega t)\right]. 
%%%
\end{eqnarray}
\begin{eqnarray}
j_{y,\alpha}&=&-e\left[\frac{\partial \alpha_k}{\partial k} \sin\phi-\frac{\Lambda}{8}(M_{k,+})^2\frac{1}{k}\mu\tau\sin\theta\cos\omega t-\frac{\Lambda}{8}\frac{1}{k}\tau\sin^3\theta(\sin\omega t \sin 2\phi+\mu\cos 2 \phi\cos\omega t)\right], \\
j_{y,\beta}&=&-e\left[\alpha_k\sin\theta\frac{1}{k}\cos\phi+\frac{\Lambda}{8}\frac{1}{k}\mu\sin^2\theta M_{k,-}\sin\omega t+\frac{\Lambda}{8}\frac{1}{k}\sin^2\theta M_{k,+}(-\cos\omega t \sin 2 \phi+\mu\cos 2 \phi\sin\omega t)\right], \\
j_{y,\gamma}&=&-e\left[-\tau\alpha_k\sin\theta\cos\theta\frac{1}{k}\sin\phi-\frac{\Lambda}{8}\mu\frac{1}{k}\sin^2\theta M_{k,-}\cos\omega t-\frac{\Lambda}{8}\frac{1}{k}\sin^2\theta M_{k,+}(\mu\cos\omega t \cos 2 \phi+\sin 2 \phi\sin\omega t)\right].
\end{eqnarray}

\subsection{First-order density matrix} \label{sec:app1st}

The solutions obtained by solving equation \eqref{matrixBCD} %, after the simplification scheme explained in Sec. \ref{sec:dcI},
are presented as follows. First, in the current expressions Eqs.~\eqref{JxtauM}-\eqref{JytauM}, we observe the following $\phi$-dependence: $\tilde{S}_{k,\alpha}^{(1)}$ is multiplied by a $\cos\phi$ or $\sin\phi$, while $\tilde{S}_{k,\beta}^{(1)}$ and  $\tilde{S}_{k,\gamma}^{(1)}$ are multipled by $1$, $\cos2\phi$ or $\sin2\phi$. Therefore, we only need to keep terms dependent on $\cos\phi, \sin\phi$ in $\tilde{S}_{k,\alpha}^{(1)}$ and terms on $1,\cos2\phi,\sin2\phi$ in $\tilde{S}_{k,\beta}^{(1)}$ and  $\tilde{S}_{k,\gamma}^{(1)}$; other terms will vanish upon integration over $\phi$. Hence we show only the relevant terms that will give non-vanishing contribution to the time-averaged longitudinal and Hall currents:

\begin{eqnarray}
 \tilde{S}_{k,\alpha}^{(1)}&=&\frac{1}{8k}\frac{eE}{D}\left\{4\mu\tau\Lambda\sin^2\theta_k\text{Im}\left\{S^T_k B^L_k\right\}\sin\phi+\left[4k\Lambda M_{k,+}  \text{Re}\left\{\frac{\partial S^T_k}{\partial k} B^L_k\right\}+8k\frac{\partial S_{k,0}}{\partial k}\lvert B_k^L\rvert^2\right]\cos\phi\right\}\,\cdots, \label{S1alpha} \\ 
 %%%
 \tilde{S}_{k,\beta}^{(1)}&=&\frac{1}{8k}\frac{eE}{D}\bigg\{\Big[4\tau\Gamma M_{k,-}\text{Im}\left\{S^T_k B^L_k\right\}+4\mu k \Gamma \text{Im}\left\{\frac{\partial S^T_k}{\partial k} B^L_k\right\}+2k(2\alpha_k-\omega)M_{k,+}\Lambda\frac{\partial S_{k,0}}{\partial k}+k M_{k,+}^2\Lambda^2\frac{\partial S_{k,1}}{\partial k}\Big] \nonumber \\
 &&+\Big[-4\tau\Gamma M_{k,-}\text{Re}\left\{S^T_k B^L_k\right\}+4\mu\Gamma k \text{Re}\left\{\frac{\partial S^T_k}{\partial k} B^L_k\right\}-2\mu k M_{k,+}\Gamma_\perp\Lambda\frac{\partial S_{k,0}}{\partial k}-\mu\tau\sin^2\theta_kM_{k,+}\Lambda^2 S_{k,2}\Big]\sin 2\phi \nonumber \\ 
 &&+\Big[-4\tau\Gamma M_{k,-}\text{Im}\left\{S^T_k B^L_k\right\}+4\mu k \Gamma \text{Im}\left\{\frac{\partial S^T_k}{\partial k} B^L_k\right\}+2k(2\alpha_k-\omega)M_{k,+}\Lambda\frac{\partial S_{k,0}}{\partial k}+k M_{k,+}^2\Lambda^2\frac{\partial S_{k,1}}{\partial k}\Big]\cos 2\phi\bigg\}\,\cdots,\nonumber  \\  \label{S1beta} \\
 %%%%
\tilde{S}_{k,\gamma}^{(1)}&=&\frac{1}{8k}\frac{eE}{D}\bigg\{\Big[4\mu\tau\Gamma M_{k,-}\text{Re}\left\{S^T_k B^L_k\right\}+4 k \Gamma \text{Im}\left\{\frac{\partial S^T_k}{\partial k} B^L_k\right\}-2kM_{k,+}\Gamma_\perp\Lambda\frac{\partial S_{k,0}}{\partial k}+\tau \sin^2\theta_k M_{k,+}\Lambda^2 S_{k,2}\Big] \nonumber \\
 &&+\Big[-4\mu\tau\Gamma M_{k,-}\text{Re}\left\{S^T_k B^L_k\right\}+4\Gamma k \text{Re}\left\{\frac{\partial S^T_k}{\partial k} B^L_k\right\}-2 k M_{k,+}\Gamma_\perp\Lambda\frac{\partial S_{k,0}}{\partial k}-\tau\sin^2\theta_kM_{k,+}\Lambda^2 S_{k,2}\Big]\cos 2\phi \nonumber \\
 &&+\Big[4\mu\tau\Gamma M_{k,-}\text{Im}\left\{S^T_k B^L_k\right\}-4 k \Gamma \text{Im}\left\{\frac{\partial S^T_k}{\partial k}B^L_k\right\}-2\mu k(2\alpha_k-\omega)M_{k,+}\Lambda\frac{\partial S_{k,0}}{\partial k}-\mu k M_{k,+}^2\Lambda^2\frac{\partial S_{k,1}}{\partial k}\Big]\sin 2\phi\bigg\}\,\cdots, \nonumber \\ \label{S1gamma}
\end{eqnarray}
where $S_{k,0}, S_{k,1}, S_{k,2}$ are given in Eqs.~\eqref{Sk123}, $M_{k,\pm}$ is defined under Eq.~\eqref{JytauM}, and 
\begin{eqnarray}
S_k^T&=&S_{k,1}-iS_{k,2}=-\frac{\Lambda}{2}M_{k,+}\frac{(2
\alpha_k-\omega)+i\mu\Gamma_\perp}{(2\alpha_k-\omega)^2+\Gamma_\perp^2+\left(\Lambda M_{k,+}/2\right)^2\left({\Gamma_\perp}/{\Gamma}\right)}, \\
B_k^L&=&(2\alpha_k-\omega)+i\mu\Gamma_\perp,
\end{eqnarray}
and $D$ is  the determinant of the $3\times3$ matrix in Eq. \eqref{matrixBCD},
\begin{eqnarray}
 D&=&\Gamma\left[\Gamma_\perp^2+(2\alpha_k-\omega)^2+\frac{\Gamma_\perp}{\Gamma}\left(\frac{\Lambda}{2}\right)^2M_{k,+}^2\right].  
\end{eqnarray}

\subsection{$\tilde{S}_{k,\alpha}^{(1)}, \tilde{S}_{k,\beta}^{(1)},\tilde{S}_{k,\gamma}^{(1)}$ contributions in the longitudinal and Hall conductivities}

In the following plots, we display the contributions due to $\tilde{S}_{k,\alpha}^{(1)}, \tilde{S}_{k,\beta}^{(1)},\tilde{S}_{k,\gamma}^{(1)}$ in the longitudinal [Eq.~\eqref{JxtauM}] and Hall conductivities [Eq.~\eqref{JytauM}], which supplement our discussions on our results in Fig.~\ref{Jyt1t2}.
\begin{figure}[htb]
  \includegraphics[width=18cm,angle=0]{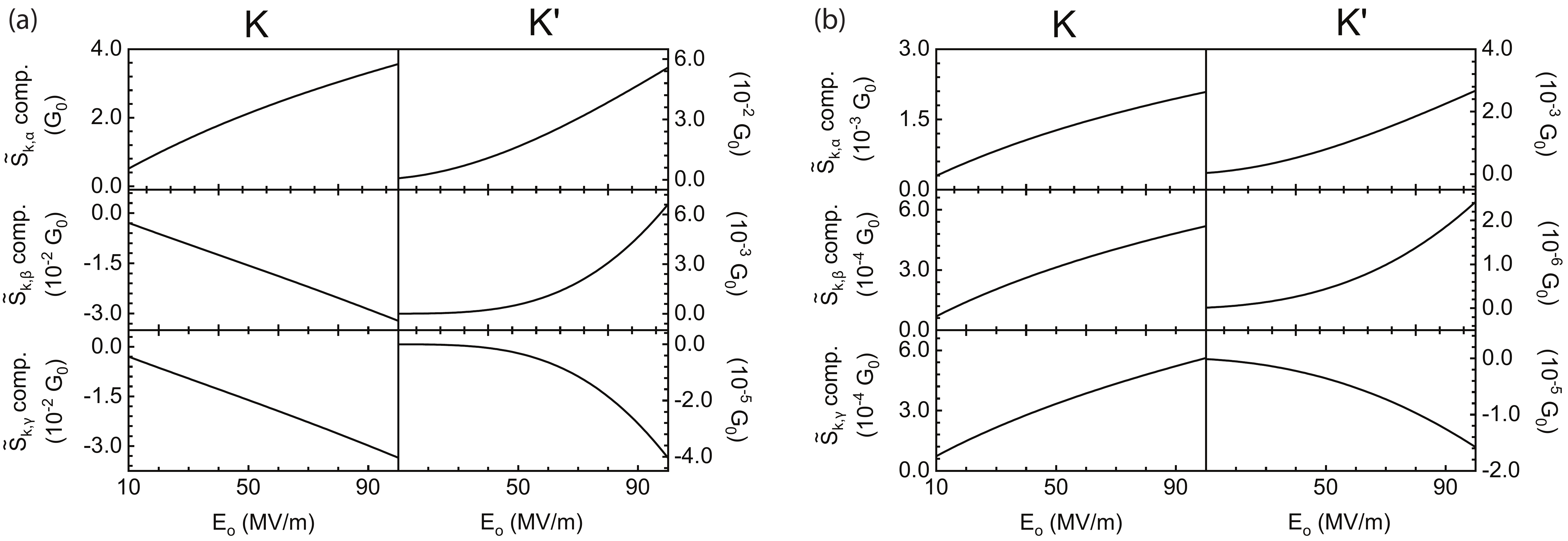}
  \caption{(Color online). Valley-specific conductivities (a) $\sigma_{xx}$ and (b) $\sigma_{yx}$ in units of $G_0 = e^2/\hbar$ as a function of the field strength $E_0$ for a pump field with frequency $\omega = 1.62\,\mathrm{eV}$ and helicity  $\mu = 1$. For each of $\sigma_{xx}$ and  $\sigma_{yx}$, the the contribution from $\tilde{S}_{k,\alpha}^{(1)}$ is  shown in the first row, $\tilde{S}_{k,\beta}^{(1)}$ in the second row and $\tilde{S}_{k,\gamma}^{(1)}$ in the third row, whereas the two columns show the cases  for valleys K and K'. Relaxation and dephasing rates are taken as $\Gamma = \Gamma_{\perp} = 1\,\mathrm{meV}$. } \label{condxx_xy_E0}
\end{figure}
\end{widetext}

\bibliography{PAHE_2_5}

\end{document}